\documentclass[manuscript,screen,sigconf]{acmart}
\AtBeginDocument{%
  \providecommand\BibTeX{{%
    \normalfont B\kern-0.5em{\scshape i\kern-0.25em b}\kern-0.8em\TeX}}}

\acmConference[ICER 2024]{}{August 13--15, 2024}{ Melbourne, Victoria, Australia}
\copyrightyear{2024}
\acmYear{2024}
\setcopyright{acmlicensed}\acmConference[ICER '24 Vol. 1]{ACM Conference on International Computing Education Research V.1}{August 13--15, 2024}{Melbourne, VIC, Australia}
\acmBooktitle{ACM Conference on International Computing Education Research V.1 (ICER '24 Vol. 1), August 13--15, 2024, Melbourne, VIC, Australia}
\acmDOI{10.1145/3632620.3671121}
\acmISBN{979-8-4007-0475-8/24/08}

\usepackage{enumitem}
\usepackage{hyperref}
\usepackage{comment}
\usepackage{mdframed}

\usepackage{graphicx}
\usepackage{caption}
\usepackage{subcaption}
\usepackage{fancyvrb} 
\usepackage{xcolor}
\usepackage{pifont}%
\newcommand{\cmark}{\ding{51}}%
\newcommand{\xmark}{\ding{55}}%

\begin{document}

\title[Influence of Personality Traits on Plagiarism Through Collusion in Programming Assignments]{Influence of Personality Traits on Plagiarism Through Collusion in Programming Assignments}

 
\author{P D Parthasarathy}\email{p20210042@goa.bits-pilani.ac.in}\orcid{0000-0002-8723-2407}
\affiliation{%
\institution{Department of Computer Science and Information Systems \\ BITS Pilani, KK Birla Goa Campus, Goa, India}
\city{}\state{}
\country{}
\postcode{}
}

\author{Ishaan Kapoor}\email{f20212091@goa.bits-pilani.ac.in}\orcid{0009-0002-4221-638X}
\affiliation{%
\institution{Department of Computer Science and Information Systems \\ BITS Pilani, KK Birla Goa Campus, Goa, India}
\city{}\state{}
\country{}
\postcode{}
}

\author{Swaroop Joshi}\email{swaroopj@goa.bits-pilani.ac.in}\orcid{0000-0003-4536-2446}
\affiliation{%
\institution{Department of Computer Science and Information Systems \\ BITS Pilani, KK Birla Goa Campus, Goa, India}
\city{}\state{}
\country{}
\postcode{403726}
}

\author{Sujith Thomas}\email{sujitht@goa.bits-pilani.ac.in}\orcid{0000-0001-6955-3664}
\affiliation{%
\institution{Department of Computer Science and Information Systems, APPCAIR, BITS Pilani, KK Birla Goa Campus, Goa, India}
\city{}\state{}
\country{}
\postcode{403726}
}

\renewcommand{\shortauthors}{}

\begin{abstract}
    
Educating students about academic integrity expectations has been suggested as one of the ways to reduce malpractice in take-home programming assignments. We test this hypothesis using data collected from an artificial intelligence course with 105 participants (N=105) at a university in India. The AI course had two programming assignments. Plagiarism through collusion was quantified using the Measure of Software Similarity (MOSS) tool. Students were educated about what constitutes academic dishonesty and were required to take an honor pledge before the start of the second take-home programming assignment. The two programming assignments were novel and did not have solutions available on the internet. We expected the mean percentage of similar lines of code to be significantly less in the second programming assignment. However, our results show no significant difference in the mean percentage of similar lines of code across the two programming assignments. We also study how the Big-five personality traits affect the propensity for plagiarism in the two take-home assignments. Our results across both assignments show that the extraversion trait of the Big Five personality exhibits a positive association, and the conscientiousness trait exhibits a negative association with plagiarism tendencies. 
Our result suggests that the policy of educating students about academic integrity will have a limited impact as long as students perceive an opportunity for plagiarism to be present. We explain our results using the Fraud triangle model.

\end{abstract}




\newenvironment{myquote}%
  {\list{}{\leftmargin=0.1in\rightmargin=0.1in}\item[]}%
  {\endlist}

\maketitle

\section{Introduction} \label{sec:intro}

Academic integrity, essential for upholding ethical standards in academia, is often gauged by behaviors that defy it—such as seeking and using unauthorized assistance in assignments and exams. It remains a pivotal challenge in education. Computer Science~(CS) students often face challenges in completing programming assessments, stemming from poor time management and a scarcity of suitable resources \cite{10.5555/858403.858418}. When students cannot access sufficient support, they may resort to cheating as a desperate measure \cite{10.1145/3059009.3059065, 10.1145/1632149.1632168}. Over the years, a wide range of research has investigated this topic, from exploring why students plagiarize to developing automatic methods of detecting those who do \cite{10.5120/ijca2015906113}.

Researchers have identified various forms of academic dishonesty \cite{40db574cf24f441d994cbb1cd909bd1f}, including but not limited to:

\begin{itemize}
    \item Collaborating excessively on take-home assignments,
    \item Copying assignments in part or whole from other students,
    \item Seeking assistance from the Internet to solve challenging problems,
    \item Submitting identical work for multiple courses,
    \item Plagiarizing text from external sources,
    \item Outsourcing the assignment to someone by paying them,
    \item Using undisclosed resources during examinations, among other practices.
\end{itemize} 

Academic dishonesty is classified into three primary categories: cheating, plagiarism, and collusion \cite{8937362}. In this work, we focus on the latter two. While some use the term `plagiarism' to encompass both, a distinction exists between the two in the context of CS disciplines. \textit{Plagiarism} involves using the work of `others' without proper attribution, typically sourced from public mediums like GitHub, journals, or the internet. On the other hand, \textit{collusion} involves collaborating with `known' individuals (typically classmates) to produce academic work without proper attribution. An example of plagiarism would be incorporating unattributed content from the web, while collusion involves students collaborating on individual assignments meant to be completed independently \cite{Owunwanne, Fraser2014CollaborationCA,JonesJuliet}.

While much of the existing research on academic integrity focuses on textual plagiarism, there is growing recognition that plagiarism can occur across various non-text mediums such as source code and multimedia \cite{Marcin2016,10.1145/2526968.2526971}. According to McCabe et al.\cite{84f6c08b47844b54868caf82c625fc66}, behaviors such as cut-and-paste, which faculty members view as plagiarism, are becoming more acceptable to students. Even when students recognize its wrongness, many admit to engaging in academic dishonesty during their college years\cite{Bernardi2004-BERETD}. Addressing academic integrity within the computing domain poses distinct challenges due to inherent issues within the discipline. The usual rules and policies of the university do not always fit well with programming assignments. For instance, attributing source code according to standard practices such as quoting the original work or citing them may compromise the syntax of the code and can lead to syntax errors. Source code attribution is not commonly done using comments because it can reduce readability, maintainability, and focus. In undergraduate programming assignments, code stubs or boilerplate code are often provided, and students are encouraged to use examples used in the class. Since most students use these provided code stubs and examples, the suspicion of plagiarism increases \cite{10.5555/1151869.1151888}. 
Additionally, in programming assignments involving low level programming (e.g. assembly language) there may be limited ways to express certain features \cite{10.1145/3160489.3160502}, increasing the suspicion of plagiarism. 
Despite ongoing efforts, a universally accepted format for attributing externally sourced program code remains elusive \cite{10.1145/1562877.1562900}. The ITiCSE’16 Working Group \cite{10.1145/3024906.3024910} advocated for instructors to articulate their expectations regarding academic integrity in their courses clearly and provided examples of how to convey them. Simon et al.\cite{10.1145/3160489.3160502} subsequently suggested a standardized form for acknowledging the use of source code authored by others. 

While most students clearly recognize behaviors like exam fraud and outsourcing assignments as forms of cheating, there is notable confusion among both students and academics within the computing field regarding the distinctions surrounding plagiarism and collusion in programming assignments \cite{SCPlagiarism}. Numerous studies have highlighted genuine misunderstandings among students regarding what constitutes plagiarism in programming assignments \cite{SCPlagiarism, Cosma2008TowardsAD, 10.1145/637610.544468, 10.1145/2526968.2526971}. This lack of clarity among both faculty and students heightens the risk of unintentional and possibly inadvertent engagement in inappropriate academic practices. 

Multiple researchers suggest educating students about plagiarism (both coincidental and non-coincidental), and establishing clear academic integrity standards as a possible approach to mitigating instances of plagiarism and collusion in take-home programming assignments\cite{10.1145/3506717, 10.1145/2591708.2591755, 10.1145/2632320.2632342, Joy2011SourceCP}. However, to the best of our knowledge, this hypothesis has not been extensively tested developing countries like India. Given the substantial enrollment of over 2 million undergraduates enrolled in computer science and related courses across Indian higher education institutes \cite{departmentofhighereducationgovernmentofindiaAllIndiaSurvey2021}, along with raising concerns on student plagiarism within the country\cite{bakthavatchaalam2021academic} and among Indian students studying outside India\cite{plagiairmInIndia}, it becomes crucial to evaluate the efficacy of these strategies in varied cultural and geographical landscapes.

In our study, we clearly outline to students the guidelines regarding plagiarism and collusion in programming tasks, along with obtaining signed honor pledges. We then examine how these measures affect students' conduct concerning plagiarism and collusion.

On a parallel trajectory, the Big Five personality traits, also known as the Five Factor model or OCEAN model (explained further in the next section), are five broad dimensions that capture different aspects of human personality such as Openness to experience (O), Conscientiousness (C), Extraversion (E), Agreeableness (A), and Neuroticism (N). The influence of the Big Five personality traits is a subject of study across various fields, including its influence on anxiety \cite{AnxityBigfive}, personality development \cite{Tetzner}, and academic achievement \cite{AchievementBigfive, https://doi.org/10.1111/jopy.12663, OZ2015PER}. 

It has also been studied in the context of academic plagiarism in education \cite{Giluk2015BigFP, Bhutto2019ACS}. However, while previous research has explored the relationship between these traits and the tendency to plagiarize, much of this work focuses on predicting plagiarism propensity rather than directly measuring plagiarism scores and their correlation with personality traits. In our study, we aim to address this gap by examining the actual plagiarism scores alongside the big five trait scores of students to identify the influence of these traits on plagiarism in programming assignments. In summary, the research questions of this work are as follows:

\begin{itemize}
    \item \label{RQ1} \textbf{RQ1:} What influence do the big-five personality traits have on plagiarism in programming assignments?
    \item \label{RQ2} \textbf{RQ2:} To what extent does sensitizing students about academic integrity and the criteria for plagiarism and collusion through an honor pledge influence their behavior?
\end{itemize}

We approached RQ1 as exploratory, without a specific hypothesis. For RQ2, we anticipated a decrease in the mean percentage of similar lines of code in the second programming assignment following student sensitization through an honor pledge. This study, to our knowledge, is the first of its kind, and its findings could be valuable for the computing education community. 

The Sec \ref{sec:frameworks} briefly explains the various theoretical frameworks used in this work and relates to computing education literature in Sec \ref{sec:relatedwork}. The methodology is explained in Sec \ref{sec:method}. The results are presented in Sec \ref{sec:findings} and discussed in \ref{sec:discuss}. The limitations and next steps are highlighted in Sec \ref{sec:limitations} and our findings are concluded in Sec \ref{sec:conclusion}.

\section{THEORETICAL FRAMEWORK} \label{sec:frameworks}
This work adopts theoretical framings from personality psychology, criminology, and education. This section briefly explains the concepts from these theoretical frameworks for the readers' benefit (In the next section, we explain how they relate to computing education and this work in particular). 

\subsection{Personality Psychology}
The Big Five personality traits, the Five Factor Model (FFM), represent a widely accepted framework for understanding and categorizing human personality. This model emerged due to extensive psychological research to identify the fundamental dimensions underlying individual personality differences. The development of the Big Five model began in the late 20th century, with early research conducted by psychologists such as Tupes and Christal in the 1960s \cite{tupesRECURRENTPERSONALITYFACTORS}, followed by the work of Goldberg in the 1980s and 1990s \cite{Goldberg1990, Goldberg}, which laid the groundwork for the modern understanding of these traits.

The Big Five traits encompass five broad dimensions, each representing a distinct aspect of personality:

\begin{enumerate}
    \item \textbf{Openness to Experience (O):} This trait reflects one's inclination towards intellectual curiosity, creativity, and openness to new ideas, experiences, and perspectives. Individuals high in openness tend to be imaginative, curious, and open-minded, while those low in openness may be more conventional, cautious, and resistant to change.
    \item \textbf{Conscientiousness (C):} Conscientiousness pertains to the degree of organization, responsibility, diligence, and self-discipline an individual exhibits. High conscientiousness is associated with reliability, thoroughness, and goal-directed behavior, whereas low conscientiousness may manifest as impulsivity, disorganization, and a lack of follow-through.
    \item \textbf{Extraversion (E):} Extraversion refers to the extent to which an individual is outgoing, sociable, assertive, and energetic in social interactions. High extraversion is characterized by traits such as sociability, talkativeness, and enthusiasm, while introversion is marked by a preference for solitude, introspection, and quieter activities.
    \item \textbf{Agreeableness (A):} Agreeableness encompasses kindness, empathy, cooperativeness, and compassion towards others. Individuals high in agreeableness tend to be altruistic, trusting, and accommodating, whereas those low in agreeableness may exhibit more competitive, skeptical, or antagonistic tendencies.
    \item \textbf{Neuroticism (N):} Neuroticism reflects the tendency to experience negative emotions such as anxiety, depression, irritability, and vulnerability to stress. High neuroticism is associated with traits such as emotional volatility, insecurity, and sensitivity to perceived threats, while low neuroticism is characterized by emotional resilience, calmness, and emotional stability.
\end{enumerate}

While certain researchers have raised concerns about the Big Five model, suggesting that its emphasis on just five overarching dimensions offers an incomplete depiction of personality traits \cite{Block1995ACV}, it remains extensively embraced and applied in psychological research. Moreover, modifications have been made to the number of questions utilized to derive the scores for these five dimensions  \cite{JOHNSON201478} making it more trustworthy and accurate. 

\subsection{Criminology}
The fraud triangle, a seminal concept in criminology, was introduced by sociologist and criminologist Donald R. Cressey in 1953 \cite{Cressey}. This framework suggests that three key factors - \textit{opportunity, motivation (or pressure), and rationalization} are typically present in occupational fraud or embezzlement. According to Cressey, individuals are more likely to engage in fraudulent behavior when they perceive an opportunity to commit the act, experience pressure or motivation to do so (often due to financial difficulties or other personal stressors), and can justify or rationalize their actions. 

The comparison between fraud and plagiarism, as well as the potential use of the Fraud Triangle concept in educational settings, has been already explored in existing literature \cite{fraudTriangleToAcadmics, FraudTriEffect}. It's important to note that such discussions don't equate students with criminals but rather examine parallels in behavior and circumstances.


\section{Related Work} \label{sec:relatedwork}
In this section, we explore the existing literature and the application of the frameworks introduced in the previous section. \vspace{-0.2cm}

\subsection{The Big Five Personality Traits in Academic Plagiarism}

Interest in the influence of the big five personality traits on academic plagiarism has grown in recent years. However, research in this area remains limited within computing education. Contradictory findings exist in the literature; for example, Bhutto et al. \cite{Bhutto2019ACS} explored the correlation between personality traits and plagiarism among 231 undergraduate-level students from social science departments, revealing \textit{positive} associations with agreeableness, conscientiousness, extraversion, and openness to experience, while no significant relationship was found with neuroticism. Conversely, Correa \cite{ChileanUniver} investigated cyber plagiarism among 106 Chilean undergraduate students across departments, finding a significant \textit{negative} correlation with conscientiousness, extraversion, and openness to experience and a positive correlation with neuroticism. Both of their works use a questionnaire measuring personality traits and self-reported plagiarism behavior.

Wilks et al. \cite{Wilks2016-WILPTA-3} conducted the same study in the Portuguese context with undergraduate students from Law and Criminology departments and found that Conscientiousness and Agreeableness traits are negatively correlated with the inclination to plagiarize, while no significant association was found with Neuroticism. 

Karim et al. \cite{karimExploringRelationshipInternet2009} explored the extent of unethical behavior such as fraudulence, plagiarism, unauthorized help, downloads, and copy-paste functions at a public university in Malaysia and its relationship with the Big Five personality traits. They conducted the study by surveying 252 students from across three universities. The students were chosen from different majors, such as Economics, Information and Communication Technology (ICT), human science, and engineering. They found that agreeableness, conscientiousness, and neuroticism are significantly and negatively correlated with unethical internet behavior. Additionally, they found significant differences in plagiarism between the universities. Their findings highlight the importance of incorporating computer ethics into the curriculum and developing policies to guide academic conduct and Internet use.

Giluk et al.\cite{Giluk2015BigFP} performed a meta-analysis to estimate the relationship between each of the Big Five personality factors and academic dishonesty. Their findings indicate that conscientiousness and agreeableness are the strongest Big Five predictors, with both factors negatively related to academic dishonesty. Table \ref{tab:litSUmmary} summarizes the literature regarding the impact of each of the big five traits on academic plagiarism. A positive (+ve) or negative (-ve) sign denotes the direction of influence, while "-" indicates no significant effect. "NA" indicates not applicable (as it was a meta-analysis). \vspace{-4pt}

\begin{table}[H]
  \centering
  \caption{Summary of Big Five traits and Plagiarism\\\label{tab:litSUmmary}}
    \vspace{-12pt}
  \begin{tabular}{p{2cm}rrrrrr}
    \toprule
    Study & Participants & O & C & E & A & N \\\midrule
    Bhutto et al.\cite{Bhutto2019ACS} & 231 & +ve & +ve & +ve & +ve & - \\
    Correa\cite{ChileanUniver} & 106 & -ve & -ve & -ve & - & +ve \\
    Wilks et al.\cite{Wilks2016-WILPTA-3} & 373 & - & -ve & - & -ve & - \\
    Karim et al. \cite{karimExploringRelationshipInternet2009} & 252 & - & -ve & - & -ve & -ve \\
    Giluk et al.\cite{Giluk2015BigFP} & NA & - & -ve & - & -ve & - \\ \bottomrule
  \end{tabular} \vspace{-8pt}
\end{table}

The findings suggest inconclusiveness, necessitating further replication and substantiation through additional evidence. Moreover, it's crucial to recognize that these studies depend on self-reported occurrences of plagiarism or student collusion, which may not offer a precise representation due to the potential influence of fear regarding disciplinary consequences or social desirability bias. 

Notably, there exists a significant dearth of such research within computing education. Our study aims to fill this gap by addressing the inquiry outlined in our research question RQ1. Additionally, rather than relying on self-reported plagiarism behavior by students, we assess actual assignments to determine collusion and plagiarism scores.

\subsection{Plagiarism in Programming Assignments}
Plagiarism in computing education has been researched heavily in the past two decades. Naaj et al. \cite{8937362} explored the factors impacting plagiarism and collusion in programming assignments, revealing that female students and those with a higher CGPA tended to exhibit stronger ethical behavior compared to their peers. Additionally, no significant differences were observed based on student level or major. 

Karnalim et al. \cite{9678917} conducted a study in Indonesia aimed at understanding student perceptions of programming plagiarism and collusion. Their survey of 345 students from 16 universities revealed that students generally recognize instances of programming plagiarism and collusion, except for behaviors like self-plagiarism, copying from early drafts of their own work, and seeking assistance to correct code. A more recent study by Liut et al. \cite{10.1145/3626252.3630753} investigates what students understand about academic integrity in computer science~(CS) courses and if there are differences based on university, country, demographic factors, or online versus in-person courses. They surveyed 1,011 undergraduate CS students at three universities (Australia, Canada, and the USA). Their findings show that all three institutions take academic integrity seriously, and students recognize its importance, yet confusion about policy specifics is common. Their findings also suggest that course instructors significantly influence what students perceive as violations of academic integrity policies.

Karnalim et al \cite{10.1145/3506717} create and evaluate an assessment submission system with automated, personalized feedback to deter collusion and plagiarism. The system generates personalized similarity reports for submissions showing similarities, prompting students to clarify and resubmit their work. Otherwise, similarity reports are sent only to the submitting student to improve their understanding. Their research suggests decreased programming plagiarism and collusion following the system's implementation, although student engagement was lower. To address this, gamification elements were introduced, offering points and badges for reviewing generated reports. Those with the highest points were offered rewards. As a result, students demonstrated increased awareness of programming plagiarism and collusion, completed assessments sooner, and engaged more frequently with the reports \cite{10042043}.

When comparing student programs for similarity to detect academic misconduct, certain code segments, such as boilerplate code (e.g., \Verb#public static void main String[] args#) and code provided in assignments, are often similar but not suspicious. An ITiCSE'20 working group collected and analyzed assessment submissions from various institutions to identify common code segments and their nature and rationale. Their first research question examined which common code segments should be excluded from similarity detection. They identified types of common segments but not indicative of plagiarism or collusion, explaining their prevalence and typical assignment contexts. Their second research question explored the impact of removing common code segments on similarity detection reports. Excluding specified common segments from many detection tools resulted in fewer and shorter matched segments. Their proposed approach enhances the effectiveness of similarity detection by focusing on truly suspicious matches and improves manual checking efficiency by reducing the number of segments to review.

Honor pledges have demonstrated effectiveness in educating students about plagiarism and collusion. Gibbs et al. \cite{10.1145/2724660.2728663} found that while honor codes had a non-significant effect, pre-task warnings significantly reduced cheating rates. Bates et al.\cite{bates2016academic} conducted experiments using two approaches to educate students about plagiarism: providing a student code of conduct and presenting explicit definitions of plagiarism from their university's senate rules. Both interventions resulted in shifts in students' perceptions of acceptable academic behavior, underscoring the importance of clearly defining boundaries for unacceptable conduct. 

Similarly, in the summer of 2017, Mason et al. \cite{10.1145/3287324.3287443} modified the delivery of honor pledges for Masters students in a CS program. They integrated a clear explanation of university and class policies on academic honesty with a short formal assessment of students' understanding of plagiarism expectations through their LMS. This approach led to a significant reduction in plagiarism rates. They repeated the experiment for over four semesters and demonstrated that the approach was consistently effective.

Albluwi \cite{10.1145/3371156} conducted a comprehensive systematic review of plagiarism in programming assignments in 2019 with 87 published papers from the lens of the fraud triangle. The review revealed that a majority (68\%) of the examined papers focused on methods to reduce \textit{opportunities} for plagiarism and tools for detecting it. However, there is a notable absence of empirical research assessing the effectiveness of these strategies and tools as deterrents. Additionally, the papers (33\%) discussed various \textit{rationalizations} employed by computing students to justify plagiarism, with genuine confusion about plagiarism definitions being a prominent factor. Additionally, research on the correlation between academic \textit{pressure} in computing courses and plagiarism was limited.

Research studies have consistently highlighted significant confusion among students regarding the distinction between acceptable collaboration and unacceptable collusion \cite{10.1145/2591708.2591755, 10.1145/2632320.2632342, Joy2011SourceCP}. While researchers have underscored the importance of clearly communicating what constitutes plagiarism to students \cite{10.1145/1595496.1562900, 10.1145/3160489.3160502, 10.1145/1734263.1734365, 10.1145/3024906.3024910}, examining the effectiveness of such efforts is limited. This study addresses this gap through research question RQ2.

\section{Methodology} \label{sec:method}

In this section, we provide a detailed overview of the context in which our study was conducted, including the setting, participant details, measurement tools employed, and the research methodology adopted.

\subsection{Setting and Participants}
The research was conducted within an Artificial Intelligence (AI) elective course, consisting of three credits. In our institute, three credits correspond to a total of nine hours of work during a week, including three 1 hour lectures. The AI course was offered during the Fall semester of 2023 at a large private university in India. The study involved 105 junior and senior undergraduate students pursuing computer science and information systems degrees. Among the participants, 98 were male, and 7 were female, all within the age range of 18 to 21 and hailing from India. The course employed a continuous evaluation method, with 30\% allocated to two programming assignments, 30\% to a mid-term closed-book examination, and 40\% to a comprehensive closed-book final examination at the end of the term.

\subsection{Measures}
The following measures were used in the study:
\subsubsection{The International Personality Item Pool (IPIP)}
The International Personality Item Pool (IPIP)\footnote{\url{https://ipip.ori.org/}} stands as a comprehensive resource for personality assessment items, facilitating research in psychology across diverse cultural settings. Established in the 1990s by Goldberg \cite{Goldberg}, the IPIP offers an extensive array of self-report measures to assess various dimensions of personality, including the Big Five traits: openness, conscientiousness, extraversion, agreeableness, and neuroticism. In this study, participants' Big Five personality traits were evaluated using the rigorous and comprehensive 120-item version of the IPIP \cite{JOHNSON201478}. This choice was made because the 120-item version strikes a balance, being neither too brief (like the 50-item version) nor overly extensive (like the 300-item version), thus enabling completion within a reasonable timeframe of less than 30 minutes, which is suitable for students.

The 120-item version of the IPIP scale comprises 24 items for each subscale, totaling 120 items. Participants assess the degree to which each item (question or statement) reflects their personality on a 5-point Likert scale, ranging from 1 = very inaccurate to 5 = very accurate. The internal consistency coefficients (Cronbach's alpha) for the five subscales ranged from \begin{math} \alpha \end{math} = 0.84 to 0.93, with values for extraversion at 0.92, agreeableness at 0.85, conscientiousness at 0.84, neuroticism at 0.90, and openness to experience at 0.93. These coefficients signify the reliability of the scale in measuring each personality trait, with higher values indicating greater internal consistency among the items within each subscale.

\subsubsection{Plagiarism Detection}

The Measure of Software Similarity (MOSS) tool \cite{MOSS} was used to detect plagiarism in the programming assignments through collusion among students. MOSS provided a percentage-based measure of similarity between code submissions, allowing for the quantification of plagiarism in each of the two programming assignments. The following command was used to get the count of similar lines: 
\begin{verbatim}
./moss.pl -l python -b ProgramBaseFile/Base_program.py \
Programs/*.py
\end{verbatim}

MOSS usually reports all code matches in pairs of program files. However, when a `base file' is provided using the \textbf{-b} option, the lines of code present in the base file are not counted in the matches reported by MOSS. This ensures that any matching code between two students is not due to the base program provided by the instructor for the programming assignment. The preference for the MOSS tool over JPlag or other source code plagiarism detection tools stemmed from the authors' familiarity with MOSS. In future studies, we intend to incorporate multiple tools to ensure comprehensive plagiarism detection.

\subsection{Research Design}

\begin{figure}[H]
  \centering
  \frame{\includegraphics[width=\columnwidth]{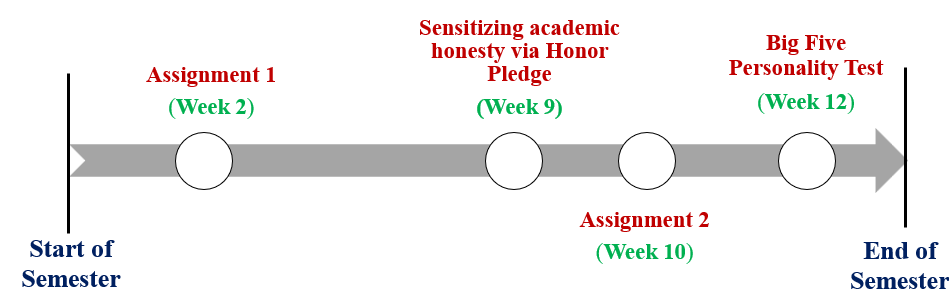}}\vspace{-8pt}
  \caption{Series of activities}
  \label{fig:activityTimeline} \vspace{-10pt}
\end{figure}

As depicted in Figure \ref{fig:activityTimeline}, the first programming assignment was given two weeks from the start of the semester. The second programming assignment was given ten weeks from the start of the semester. Owing to the significant usage of honor pledges to educate students on academic integrity expectations in literature (as discussed in Sec \ref{sec:relatedwork}), we asked students to take an honor pledge during the ninth week. The honor pledge was a pre-requisite for the second programming assignment; all 105 students participated in the pledge. 

The Big Five personality trait assessment test was done after the students submitted the second programming assignment during the twelfth week. This sequencing was deliberate to prevent students from discerning our interest in studying their behavior during the programming assignment. The first programming assignment was rolled out without explicit guidance or instructions regarding plagiarism. (However, it was mentioned in the course handout that plagiarism in programming assignments will be strictly dealt with as per the institute policy.)

 The data on personality traits and plagiarism scores, obtained through the Measure of Software Similarity (MOSS) for assignment 1, were utilized to address research question RQ\ref{RQ1}.

In the ninth week, all students had to read and endorse an honor/integrity pledge concerning plagiarism and collusion. This was aimed at raising awareness about plagiarism and upholding academic integrity. It explicitly outlined the behaviors considered plagiarism and collusion. The pledge was administered via a Google form, and a digital signature was mandated from each participant, signifying their commitment to refraining from plagiarism and collusion. Instead of in-class instruction, we opted for this method to ensure that all participants had sufficient time to comprehend the concepts of plagiarism and collusion without sacrificing any class time. The pledge remained accessible for a week and was mandatory. An excerpt from the honor pledge is provided below:

\begin{myquote}
    \textit{The assignments are to be done individually by each student. The objective of the assignments is to help students deepen their understanding of some of the concepts in AI. A secondary objective of the assignments is to help students get better at Python programming.}
    \textbf{Each} of the following amounts to violations of academic integrity:
    \begin{itemize}
        \item \textit{Sharing a part or whole of your program with another student even if one of the students changes the variable names and function names and rearranges the functions within the program file.}
        \item \textit{Sharing a part or whole of your report with another student even if one of the students changes most of the sentences so that superficially, the reports look different. You must cite the resources you have referred in the report.} 
        \item \textit{Getting parts of your program from past programming assignment submissions or internet sources. Each student must do the assignment individually.}
        \item \textit{Sharing code for commonly needed functionality by calling it "driver program" amounts to a violation of academic integrity because one of the objectives of this assignment is to help students get better at Python programming. }
    \end{itemize}
\end{myquote}
The second assignment served as the post-intervention assessment, allowing for the evaluation of the impact of plagiarism awareness on participant behavior.

Both programming assignments were original and did not have pre-existing solutions available on the internet. Below, we provide a brief overview of each assignment for the reader's understanding:

\subsubsection{First Programming Assignment:} The assignment centered on multi-armed bandit concepts and the Markov reward process. Students were tasked to develop a driver program that applied the concepts learned in class. Specifically, they were required to create a Python program capable of adjusting parameters within a provided AI model and visually representing the alterations on a graph. Additionally, students had to report all the findings along with the conclusion and rationale (as comments in the code). All relevant parameters and anticipated outcomes were covered in class, with the professor and teaching assistants available for any inquiries, during the designated assignment period via office hours. The following are the assignment's learning outcomes (LO) with their corresponding Revised Bloom Taxonomy \cite{bloomsTaxonomy} verbs in bold:
\begin{itemize}
    \item \textbf{Understand and Apply} multi-armed bandit and Markov reward process concepts. 
    \item \textbf{Analyze} parameter manipulation and visualization techniques. 
    \item \textbf{Implement} python programs efficiently 
\end{itemize}

\subsubsection{Second Programming Assignment:} The second assignment was based on the game `Connect 4'\footnote{\url{https://en.wikipedia.org/wiki/Connect_Four}}. Specifically, students were given the stub code for one version of the game that they could play and were tasked with modifying and enhancing the given program incrementally that plays against a Myopic player (a player that only views a fixed number of possibilities to evaluate a game state and not all possibilities) using the game tree-based search techniques and constraints given. Also, the students were given clear guidelines and steps to incrementally improve their player function by techniques like `Alpha-Beta Pruning'. The following are the assignment's LO's:
\begin{itemize}
    \item \textbf{Apply} game tree-based techniques and alpha-beta pruning effectively.
    \item \textbf{Implement} an incremental solution with retrospection after each stage.  
    \item \textbf{Comprehend} given python program and enhance it.
\end{itemize}

\subsection{Procedures for data collection and analysis}
\subsubsection{Personality Traits Data:} Following the institute's Human Ethical Committee (HEC) policy, participants were required to provide informed consent before undergoing the personality test. Upon obtaining consent, participants underwent the International Personality Item Pool (IPIP) personality test, which assesses the Big Five personality traits. The test was administered through an online tool available at the provided URL\footnote{\url{https://bigfive-test.com/}}.

After completing the test, participants' final scores for each Big Five personality trait were recorded in a Google form after obtaining the informed consent form as approved by the HEC. To ensure confidentiality and protect participants' privacy, all data collected was anonymized, meaning any personally identifiable information was removed or obscured.

\subsubsection{Assignment Data:} 
Both assignments were administered using Moodle, the institute's learning management system. Upon completion, the source codes for both assignments were subjected to analysis using the Measure of Software Similarity (MOSS) tool to generate plagiarism reports.

To ensure confidentiality, the data was anonymized before processing. The plagiarism scores corresponding to each student (either via collusion or direct plagiarism from other sources), derived from the MOSS reports, were then stored in cloud storage for subsequent analysis. The plagiarism score for each student was calculated based on the proportion of code similarity detected by MOSS. Specifically, it was determined by dividing the maximum number of lines copied from other sources by the total number of lines in the student's source code (ignoring the empty lines). This approach provided a quantitative measure of the extent of plagiarism in each student's assignment.

\section{Results}
\label{sec:findings}
The results of descriptive statistics revealed that 9\% of the participants rated extraversion, 14\% rated neuroticism, 17\% rated openness to experience, 28\% rated agreeableness, and 32\% rated conscientiousness as their dominant personality traits. Most participants exhibited high scores across various personality traits: 99\% scored high in Openness, indicating a proclivity for curiosity and imagination. Similarly, 91.32\% scored high in Extraversion, suggesting a preference for social interaction and enthusiasm. However, a substantial portion, 81.9\%, scored high in Neuroticism, indicating a tendency towards negative emotions and stress. Additionally, 88.57\% scored high in Agreeableness, reflecting a propensity for kindness and cooperation, while 81.1\% scored high in Conscientiousness, implying traits such as organization and responsibility. Table \ref{tab:correl} shows the pearson's correlations between the big-five personality traits. 

\begin{table}[h]
  \centering
  \caption{Correlations between the Big-Five Traits\label{tab:correl}}
  \begin{tabular}{p{3cm}rrrrr}
    \toprule
    Trait & N & E & O & A & C \\\midrule
    Neuroticism (N) & 1.00 & -0.35 & 0.00 & -0.09 & -0.48 \\
    Extraversion (E) & -0.35 & 1.00 & 0.17 & 0.05 &	0.22 \\
    Openness (O) & 0.00 & 0.17 & 1.00 &	0.15 & -0.07 \\
    Agreeableness (A) & -0.09 &	0.05 & 0.15 & 1.00 & 0.43 \\ 
    Conscientiousness (C) & -0.48 & 0.22 & -0.07 &	0.43 &	1.00  \\\bottomrule
  \end{tabular}
\end{table}

Additionally, utilizing the plagiarism scores obtained from MOSS for both assignments, students were categorized into plagiarism levels according to the criteria outlined by the University Grants Commission (UGC) \cite{UGCPlagiarism}, a statutory body overseeing higher education in India. The outcomes are detailed in table \ref{tab:levelsPlagiarism}.

\begin{table}[h]
  \centering
  \caption{Percentage of students (out of N=105) falling in each Levels of Plagiarism \label{tab:levelsPlagiarism}}
  \vspace{-12pt}
  \begin{tabular}{p{1cm}p{2cm}rr}
    \toprule
    Level & Similarity \% & Assignment 1 & Assignment 2 \\\midrule
    Level 0 & Less than 10\% & 44.76\% & 40.95\% \\
    Level 1 & 10\% to 40\% & 27.51\% & 40.00\%  \\
    Level 2 & 41\% to 60\% & 14.28\% & 7.61\% \\
    Level 3 & More than 60\% &13.33\% & 11.42\% \\ \bottomrule
  \end{tabular}\vspace{-8pt}
\end{table}

\subsection{Influence of Big-Five traits on Plagiarism}

One of the linear multiple regression assumptions is that the independent variables must not exhibit high (>5) multicollinearity. The multicollinearity between the five personality traits was tested using the variance inflation factor (VIF) and shown in Table \ref{tab:VIF}. All the VIF values were less than 5, indicating that multiple linear regression can be performed. 

\begin{table}[h]
  \centering
  \caption{Big-Five traits and their VIF\label{tab:VIF}}
    \vspace{-12pt}
  \begin{tabular}{p{2cm}r}
    \toprule
    Trait & VIF Value \\\midrule
    Neuroticism & 1.46 \\
    Extraversion & 1.20 \\
    Openness & 1.10\\
    Agreeableness & 1.33 \\ 
    Conscientiousness & 1.70  \\\bottomrule
  \end{tabular}\vspace{-4pt}
\end{table}

Multiple linear regression analyses were conducted, with the big-five personality traits serving as the independent variables and the plagiarism score as the dependent variable for both assignments. The ordinary least squares (OLS) method, a statistical technique used to estimate the relationship between independent and dependent variables, was utilized for the analysis. The dependent variable was plagiarism score and the independent variables were the personality traits. The outcomes are displayed in Tables \ref{tab:assign1Regression} and \ref{tab:assign2Regression}. 

\begin{table}[h]
  \centering
  \caption{Regression analysis on Assignment 1 with $R^2$=0.106\\\label{tab:assign1Regression}}
    \vspace{-16pt}
  \begin{tabular}{p{2.5cm}llll}
    \toprule
    Trait & $\beta$ Coef & Std Err & \textit{t} & \textit{p}\\\midrule
    Agreeableness & 0.3489 & 0.243 & 1.434 & 0.155 \\
    Openness & -0.3401 & 0.256 & -1.328 & 0.187 \\
    Neuroticism & -0.1314 & 0.197 & -0.666 & 0.507 \\
    Extraversion & 0.6078 & 0.216 & 2.809 & 0.006** \\
    Conscientiousness & -0.2667 & 0.239 & -1.114 & 0.026* \\\bottomrule
  \end{tabular}\vspace{-4pt}
\end{table}

$\beta$ Coef represents the change in the dependent variable per one-unit change in the independent variable. Std Err gauges the variability of the $\beta$ coefficient. \textit{`t'} denotes the ratio of the $\beta$ coefficient to its standard error. p-value indicates significance, with `*' for \emph{p} < 0.05 and `**' for \emph{p} < 0.01.

\begin{table}[h]
  \centering
  \caption{Regression analysis on Assignment 2 with $R^2$=0.128\\\label{tab:assign2Regression}}
    \vspace{-16pt}
  \begin{tabular}{p{2.5cm}llll}
    \toprule
    Trait & $\beta$ Coef & Std Err & \textit{t} & \textit{p}\\\midrule
    Agreeableness & 0.3291 & 0.211 & 1.561 & 0.122 \\
    Openness & -0.3750 & 0.222 & -1.691 & 0.094\\
    Neuroticism & -0.2919 & 0.171 & -1.708 &  0.091\\
    Extraversion & 0.4394 & 0.187 & 2.344 & 0.021*\\
    Conscientiousness & -0.5541 & 0.207 & -2.673 & 0.009**\\\bottomrule
  \end{tabular}\vspace{-4pt}
\end{table}

Based on the findings from both assignments, irrespective of the actual coefficient values, it is apparent that extraversion exhibits a positive association with plagiarism (\textit{p}<0.05), while conscientiousness demonstrates a negative association with plagiarism (\textit{p}<0.05). However, the effect of the remaining personality traits are not statistically significant (\textit{p} values greater than 0.05). \vspace{10pt}

\begin{mdframed}
\textbf{\textit{RQ1 Answer:}} Extraversion and conscientiousness traits influence plagiarism, with extraversion positively associated and conscientiousness negatively associated. The other three traits do not show a significant influence on plagiarism. 
\end{mdframed}\vspace{10pt}

Some of these observations align with those reported in the literature, as illustrated in Table \ref{tab:litSUmmary}. For example, the lack of significance for openness to experience and neuroticism, as well as the negative association of conscientiousness with plagiarism, are consistent with the findings of Wilks et al. \cite{Wilks2016-WILPTA-3} and Giluk et al. \cite{Giluk2015BigFP}. 

\begin{figure}[h]
  \centering
  \includegraphics[width=0.75\columnwidth]{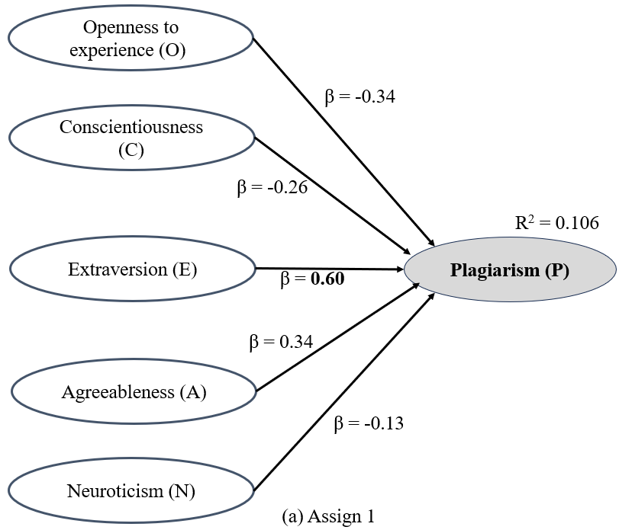}
  \includegraphics[width=0.75\columnwidth]{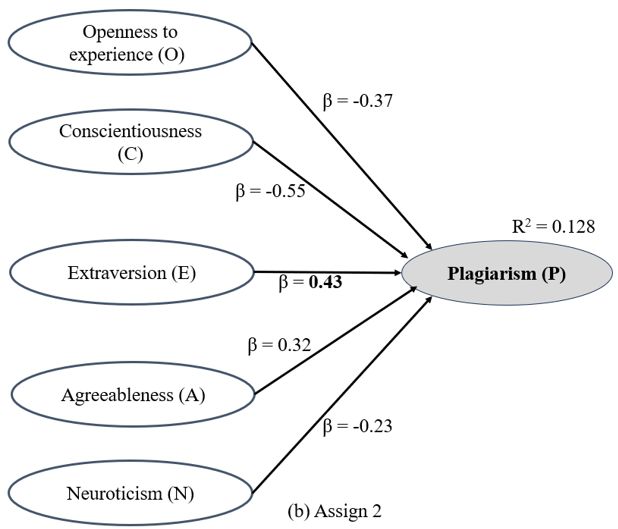}
  \caption{The relationship between the big five traits and plagiarism in (a) Assign 1 and (b) Assign 2}
  \Description{In both of them, extroversion has a positive influence and openness to experience, and conscientiousness and neuroticism negatively influence plagiarism scores.}
  \label{fig:personalityAndPlagiarism}
\end{figure}

However, in our study, extraversion demonstrates a positive association with plagiarism, consistent with the findings of Bhutto et al. \cite{Bhutto2019ACS} as shown in Fig \ref{fig:personalityAndPlagiarism}. 

\subsection{Student behavior after sensitization via Honor Pledge }

The normality of plagiarism scores for Assignment 1 and Assignment 2 was assessed using histograms and the Shapiro-Wilk test, confirming their normal distribution. Our null hypothesis suggested that there would be no change in the mean percentage of similar lines of code after students were sensitized through an honor pledge for the second programming assignment. Given that we were comparing scores from the same students pre-intervention (Assignment 1 plagiarism scores) (N=105, Mean=23.45; SD=26.25) and post-intervention (Assignment 2 plagiarism scores) (N=105, Mean=21.23; SD=23.02), we conducted a paired t-test, \textit{t}(105) = 0.89, \emph{p} = 0.374, d = 0.0899. As p>0.05, we did not reject the null hypothesis. Here, the intervention refers to the honor pledge discussed earlier (See Sec 4).

As depicted in Table \ref{tab:levelsPlagiarism}, there was a slight decrease in the percentage of students falling in levels 2 and 3 of plagiarism after the intervention. While there was a marginal decline in the average plagiarism percentage from Assignment 1 to Assignment 2 (from 23.44\% to 21.22\%), the effect size computed using Cohen's d was weak (d=0.08). \vspace{10pt}

\begin{mdframed}
\textbf{\textit{RQ2 Answer:}} Our findings suggest that while educating students about plagiarism and collusion via an honor pledge may have some effect, its effectiveness in reducing plagiarism and collusion in programming assignments is limited as long as opportunities and pressure to engage in such behavior persist.
\end{mdframed} \vspace{10pt}

We then specifically conducted the test for students who were involved in plagiarism and/or collusion in Assignment 1, those with a similarity score greater than 10\% (those falling into levels 1, 2, and 3 in Assignment 1, as shown in Table \ref{tab:levelsPlagiarism}). This allowed us to evaluate whether the intervention benefited students who engaged in such academic misconduct in Assignment 1, as they are the ones who might benefit most from such intervention. The paired sample t-test revealed a significant difference: the percentage of similarity score in Assignment 1 (N=58, Mean=41.73, SD=21.44) was notably higher than in Assignment 2 (N=58, Mean=28.92, SD=26.27); \textit{t}(57)=3.52, p=0.0008, D=0.53 (moderate effect). Similarly, when the t-test was performed only on those who did \emph{not} plagiarize and/or were involved in collusion, i.e., those whose similarity score was <10\% in assignment 1 (N=47, Mean = 0.21, SD=1.45) and assignment 2 (N=47, Mean=11.12, SD=12.56); \textit{t}(46)=5.8, p= 0.0000002, d=1.27.

While the findings regarding the significant impact of the intervention in reducing plagiarism and/or collusion on subsets of students (when considering those who were not involved separately) are compelling, it's essential to acknowledge the limitations and interpret the results with caution. Firstly, this result is not definitive as the intervention is not significant on the whole set of students (N=105) and should be further validated through replication studies, especially when considering these specific subcategories of students. Additionally, it's noteworthy that this outcome was not originally hypothesized, indicating the need for a deeper understanding of the underlying factors driving these effects.

\section{Discussion} \label{sec:discuss}

As highlighted in the Sec \ref{sec:intro}, our goal was to investigate the influence of the big-five personality traits on plagiarism and collusion in programming assignments. Additionally, we aimed to understand how sensitizing students about academic integrity and the criteria for plagiarism and collusion through an honor pledge impacts their behavior. This section examines our findings in light of the fraud triangle theory.

Our findings indicate that individuals with high extraversion scores demonstrate a greater tendency to engage in collusion to perform better in programming assignments. This propensity aligns with the fraud triangle model, where heightened extraversion fosters sociability and potentially larger social circles and friends, thereby increasing \emph{opportunities} for collusion. Moreover, extroverted individuals may find it easier to \emph{rationalize} their actions as commonplace due to their sociable nature. 

Our next finding suggests that individuals with high conscientious scores demonstrate a lesser tendency to indulge in plagiarism. Conscientious individuals resist plagiarism by naturally countering the elements of the fraud triangle model: opportunity, pressure, and rationalization. Their high personal standards and strong sense of integrity diminish the opportunity for dishonest behavior, as they are committed to ethical conduct. Effective time management and diligence reduce the pressure to resort to plagiarism due to impending deadlines or academic demands. Lastly, their intrinsic motivation for genuine learning and achievement makes it difficult for them to rationalize unethical behavior, as they prioritize the integrity of their own efforts and the long-term value of their education.

Table \ref{tab:resultCompare} presents the comparison of our findings with the findings from prior work as described in Table \ref{tab:litSUmmary}. An `\xmark' indicates that our findings do not align with the results of the specified study, while a `\cmark' signifies that our findings are consistent with their results for the respective Big-Five personality trait. Notably, the finding that extraversion positively correlates with plagiarism traits is consistent with the findings of Bhutto et al.\cite{Bhutto2019ACS}, while the negative correlation between the conscientiousness trait and plagiarism aligns with most previous studies. 

\begin{table}
  \centering
  \caption{Comparison of our findings with results from prior work as in Table \ref{tab:litSUmmary}\\\label{tab:resultCompare}}
    \vspace{-12pt}
  \begin{tabular}{p{2cm}rrrrrr}
    \toprule
    Study &  O & C & E & A & N \\\midrule
    Bhutto et al.\cite{Bhutto2019ACS} & \xmark & \xmark & \cmark & \xmark & \cmark \\
    Correa\cite{ChileanUniver} & \xmark & \cmark & \xmark  & \cmark & \xmark \\
    Wilks et al.\cite{Wilks2016-WILPTA-3} & \cmark & \cmark & \xmark  & \xmark & \cmark \\
    Karim et al. \cite{karimExploringRelationshipInternet2009} & \cmark & \cmark & \xmark  & \xmark & \xmark \\
    Giluk et al.\cite{Giluk2015BigFP} & \cmark & \cmark & \xmark  & \xmark & \cmark \\ 
    \bottomrule
  \end{tabular} \vspace{-0.80cm}
\end{table}

The correlation between the extraversion trait tending to plagiarise, making use of the opportunity, and plagiarism in India's academic environment can be understood within the \textit{pressure} aspect of the fraud triangle model. In this model, pressure refers to the internal or external forces that compel individuals to engage in misconduct. In Indian society, there is often a strong emphasis on competitiveness and achievement, particularly in academic settings where grades and academic success are highly valued \cite{IITPrepMenace,UnemploymentIndia, NAIR2020831}. Furthermore, the pressure to obtain employment is aggravated by the harsh realities of the job market. India has approximately 2 million computer science students, with ongoing concerns regarding employability rates, which can drop to as low as below 25\% \cite{UnemploymentIndia, NAIR2020831}. 

The pressure to succeed in this fiercely competitive environment and pressure stemming from societal expectations, family aspirations, and the intense competition for limited opportunities may drive individuals to resort to unethical practices like plagiarism and collusion to gain a competitive edge. The significant number (81.9\%) displaying a high neuroticism score implies a likelihood of participants experiencing stress and anxiety, highlighting the role of pressure. This environment fosters a mentality of optimizing results by any means necessary. Thus, the propensity for plagiarism is driven more by opportunity than rationale, as students feel pressured to perform. 

On the other hand, while honor pledges have been effective in significantly reducing plagiarism as shown in studies by Bates et al. \cite{bates2016academic} and Mason et al. \cite{10.1145/3287324.3287443}, our study observed a weaker impact. One possible explanation is that in prior work, the honor pledges explicitly mentioned the use of sophisticated plagiarism detection tools, whereas in our study participants were only informed about what constitutes plagiarism without being told that their submissions would be checked for plagiarism. Students perceived that there was still an \emph{opportunity} for plagiarism. Our results indicate that sensitizing students about plagiarism did not lead to a significant reduction in such behavior. Students continue to plagiarize through collusion as long as they perceive an \textit{opportunity} and/or are \textit{pressured} to do so. Consequently, the impact of the honor pledge on reducing plagiarism was perceived as weak. 


Our results suggest that individuals exhibiting higher levels of conscientiousness tend to show reduced inclinations toward plagiarism. This observation is consistent with the fraud triangle model. Conscientious individuals prioritize diligence and self-discipline, making them less prone to exploiting opportunities for plagiarism and less likely to rationalize such behavior, even under pressure.

These factors collectively imply that plagiarism and collusion in assignments may result from a blend of cultural, societal, and educational influences shaping individuals' attitudes and behaviors regarding academic integrity. Moreover, an honor pledge appears to have limited efficacy in enhancing academic integrity.

\section{Limitations and Future Work}
\label{sec:limitations}
This study has several limitations that should be noted. As a result, the findings may not be directly applicable to other educational environments or programming tasks that differ significantly in scope and complexity.

\begin{itemize}
    \item \textit{Single Institution Context:} Our study was conducted in a single institution in India, which may limit the generalizability of our findings. This study must be replicated in various educational settings and cultures to increase the generalizability. 
    \item \textit{Sample Size and Demographics:} The relatively small sample size of 105 students may not capture the diversity of student behavior and attitudes toward plagiarism. In future studies, larger, and more diverse samples will be used to validate our findings.
    \item \textit{Influence of Pre-existing Academic Integrity Culture:} The existing academic integrity culture at our institution may have influenced the outcomes. Institutions with varying levels of emphasis on academic integrity might experience different results from the intervention; thus, the results should be interpreted with caution.
    \item \textit{Impact of External Factors:} As discussed in the discussion, external factors such as peer influence, \textit{pressure} to excel, perceived \textit{opportunity}, and the difficulty of the assignments, could have contributed to the outcomes. These factors were not controlled for in the study and may have influenced the results.
    \item \textit{Absence of Control Group:} We could have designed the experiment with two groups – a control group and an honor pledge group – for each of the two programming assignments. However, we felt that the results would be difficult to interpret if collusion occurred between students of different groups. For example, if a student from the control group colluded with a student from the honor pledge group, it would complicate the interpretation of the results. Therefore, we avoided using separate groups and the current study focuses only on comparing the percentage of code similarity across two different programming assignments. Although we avoided using separate control and honor pledge groups to prevent collusion and interpretational difficulties, this decision introduces a potential confounding variable.
    \item \textit{Random Chance and Personality Traits:} The cross-sectional design of the study limits our ability to establish causal relationships between personality traits and plagiarism behavior. It is possible that, due to random chance, participants who engaged in plagiarism in both programming assignments had higher scores for the extraversion personality trait. This could result in a positive correlation between extraversion and the percentage of code similarity across both programming assignments.
    \item \textit{Gender Distribution and Cultural Homogeneity:} In India, engineering programs exhibit a severe imbalance in gender representation. For example, at our institution, only 12.5\% of the undergraduate students in the computer science program are females. The same is reflected in this study with only 6.6\% of participants being female. This disproportionate gender distribution (98 men to 7 women) and cultural homogeneity (all participants from India) may limit the generalizability of the study's findings to other demographic groups or cultural contexts.  
    \item \textit{Personality Test Timing:} Personality test was administered after the second assignment so that students won't know that their behaviour was being studied. However,  experiences during the course or external factors can alter personality trait expressions \emph{temporarily}.
\end{itemize}

Furthermore, the study primarily relied on quantitative analysis, overlooking the nuanced qualitative aspects of students' perceptions and experiences regarding plagiarism and collusion which could bring about contextual factors, such as cultural norms and academic pressures, which may influence plagiarism behavior.

Future research endeavors could address these limitations by incorporating qualitative interviews to gain deeper insights into students' attitudes, motivations, and plagiarism-related experiences. Longitudinal studies could provide a more comprehensive understanding of the dynamics between personality traits and plagiarism behavior over time, particularly following interventions aimed at sensitizing students about plagiarism. Additionally, replicating the study in diverse educational settings and cultural contexts could enhance the robustness of the findings and improve generalizability.


\section{Conclusion}
\label{sec:conclusion}

In conclusion, our study involving 105 students in India revealed significant insights into the influence of personality traits on plagiarism and the effectiveness of sensitizing students about plagiarism via an honor pledge in mitigating academic misconduct. Our findings underscored that extraversion and conscientiousness traits play pivotal roles in shaping plagiarism behavior, with extraversion positively associated and conscientiousness negatively associated with plagiarism. However, the impact of educational interventions such as honor pledges on reducing plagiarism was found to be limited. This aligns with the framework of the fraud triangle, which emphasizes the role of rationalization, opportunity, and pressure in fostering fraudulent behavior. In the highly competitive academic landscape of India, where pressure to excel is pronounced, addressing underlying pressures faced by students emerges as a critical strategy for promoting academic integrity. Therefore, our study highlights the importance of adopting comprehensive approaches that not only focus on sensitizing students about plagiarism but also address the underlying pressures contributing to academic misconduct.

\bibliographystyle{ACM-Reference-Format}
\bibliography{main.bib}



\end{document}